\documentclass[10pt,Journal]{IEEEtran}\IEEEoverridecommandlockouts
\usepackage{algorithmic}
\usepackage[linesnumbered,ruled]{algorithm2e}
\usepackage{amsfonts}
\usepackage{amsmath}
\usepackage{amssymb}
\usepackage{amsthm}
\usepackage{bm}
\usepackage{bbm}
\usepackage{booktabs}
\usepackage{caption}
\usepackage{CJK}
\usepackage{color}
\usepackage{cuted}
\usepackage{geometry}
\usepackage{graphicx}
\usepackage{indentfirst}
\usepackage[marginal]{footmisc}
\usepackage{multirow}
\usepackage{rotating}
\usepackage{subcaption}

\allowdisplaybreaks[4]

{\bgroup
 \addtolength\abovedisplayshortskip{#1}
 \addtolength\abovedisplayskip{#1}
 \addtolength\belowdisplayshortskip{#1}
 \addtolength\belowdisplayskip{#1}}
{\egroup\ignorespacesafterend}

\theoremstyle{plain}

\begin{document}

\title{\huge{Suppressing Beam Squint Effect For Near-Field Wideband Communication Through Movable Antennas}}

\author{Yanze Zhu, Qingqing Wu, Yang Liu, Qingjiang Shi, and Wen Chen \thanks{\hspace{0.35cm}Y. Zhu, Q. Wu, and W. Chen are with the Department of Electronic Engineering, Shanghai Jiao Tong University, 200240, China, email: yanzezhu@sjtu.edu.cn, qingqingwu@sjtu.edu.cn, wenchen@sjtu.edu.cn.}\thanks{\hspace{0.35cm}Y. Liu is with the School of Information and Communication Engineering, Dalian University of Technology, Dalian, China, email: yangliu\_613@dlut.edu.cn.}\thanks{\hspace{0.35cm}Q. Shi is with the School of Software Engineering, Tongji University, Shanghai, China, and also with the Shenzhen Research Institute of Big Data, Shenzhen, China, email: shiqj@tongji.edu.cn.}
\vspace{-0.2cm}}

\maketitle

\begin{abstract}
In this correspondence, we study deploying movable antenna (MA) array in a wideband multiple-input-single-output (MISO) communication system, where near-field (NF) channel model is considered. To alleviate beam squint effect, we propose to maximize the minimum analog beamforming gain across the entire wideband spectrum by appropriately adjusting MAs' positions, which is a highly challenging task. By introducing a slack variable and adopting the cutting-the-edge smoothed-gradient-descent-ascent (SGDA) method, we develop algorithms to resolve the aforementioned challenge. Numerical results verify the effectiveness of our proposed algorithms and demonstrate the benefit of utilizing MA array to mitigate beam squint effect in NF wideband system.
\end{abstract}

\begin{IEEEkeywords}
Movable antenna (MA), wideband communication system, near-field (NF), beam squint effect, smoothed-gradient-descent-ascent (SGDA).
\end{IEEEkeywords}

\section{Introduction}

Multiple-input-multiple-output (MIMO) technology is a critical technology to improve the spectral efficiency of the fifth-generation (5G) communication system and future wireless communications [1]. Very recently, a novel antenna architecture named movable antenna (MA) [2], also known as fluid antenna (FA) [3], has been proposed to further improve the beamforming capability. As opposed to the conventional fixed position antenna (FPA) array, which has its antenna elements fabricated on fixed positions, MAs can move freely within a specific area in programmable manner. This movability empowers antenna array with additional degree-of-freedom (DoF) and hence enhances its spatial multiplexing gain. MA array can be implemented by different techniques. For instance, the authors of [2] proposed a structure where MAs are connected with radio-frequency (RF)-chains and controlled by stepper motors. In [3], MA array was constructed using liquid metal technology.

Currently, a growing body of researches are exploring the potentials of MA array [4]-[7]. Specifically, the authors of [4] modeled the channel between a single-MA TX and a single-MA RX and theoretically proved that employment of MAs can improve signal-to-noise-ratio (SNR) performance. In [5], the authors considered deploying MA in a multi-user single-input-multiple-output (MU-SIMO) system and demonstrated that the use of MA can significantly lower the transmit power. The authors of [6] considered maximizing the minimal signal power of desired directions by jointly optimizing beamforming and MAs' positions. As reported in [6], utilizing MAs can remarkably improve beamforming gain and suppress interference. In [7], the authors investigated a wideband single-input-single-output (SISO) system assisted by MAs and verified that implementing MAs can boost the achievable rate based on theoretical analysis and antenna position optimization.

Although exciting progresses have been made in the existing literature, e.g., [4]-[7], deploying MA array in wideband system has not yet been thoroughly investigated, except for [7]. Note one prominent challenge to employ MIMO in wideband system is the \emph{beam squint effect} [8]-[11], i.e., analog beamforming gain degrades when subcarrier frequency deviates from the reference tone (usually the central carrier frequency). Besides, with the enlargement of antenna aperture, e.g., extremely-large (XL)-MIMO, and elevation of operating frequency, e.g., millimeter-wave (mmWave) and terahertz (THz), near-field (NF) propagation effect of electromagnetic waves will become apparent [10], [11]. However, most of the existing research, e.g., [4]-[7], are performed based on far-field (FF) channel model. Inspired by the above observations, we are motivated to investigate appropriate configuration method of MA array to effectively combat beam squint effect in NF wideband communication system. Our contributions are summarized in the following:
\begin{itemize}

\item To the best of authors’ knowledge, this is the first work studies beam squint effect in an NF wideband MA communication system.

\item To alleviate beam squint effect, we propose a max-min criterion to configure the MAs' positions, which maximizes the minimal analog beamforming gain over the entire wideband spectrum.

\item To resolve the above challenging problem, we propose two algorithms based on appropriate transformation and smoothed-gradient-descent-ascent (SGDA) method [12]. Especially, the SGDA-based solution is highly efficient.

\item Numerical results verify the effectiveness of our proposed algorithms and demonstrate the benefit of MA in alleviating beam squint effect in wideband system.

\end{itemize}

\vspace{-2mm}
\section{System Model and Problem Formulation}

As shown in Fig. 1, we consider a wideband MISO communication system where a 2-D planar MA array with $M$ antennas residing in the $yoz$-plane serves a single-antenna user. Without loss of generality, we assume that the MA array has a square shape of size $A \times A$, locates its center at the origin of the coordinate system and arranges its neighboring edges parallel with the $y$- and $z$-axis, respectively. Denote $\mathcal{M} \triangleq \{ 1, \cdots, M \}$ as the set of MAs and $\bm{p}_{m} \triangleq [0, y_{m}, z_{m}]^{\mathsf{T}}, \; m \in \mathcal{M}$, as the position of the $m$-th MA. Since all MAs should maintain within the array surface area, the restriction $\bm{p}_{m} \in \mathcal{C}_{\mathrm{MA}} \triangleq \{ (0, y, z)|y \in [-\frac{A}{2}, \frac{A}{2}], \; z \in [-\frac{A}{2}, \frac{A}{2}] \}$, for all $m \in \mathcal{M}$, should be satisfied with $\mathcal{C}_{\mathrm{MA}}$ indicating the feasible dwelling region of the MAs [4]-[7].

\begin{figure}[!t]
\centering
\includegraphics[scale=0.30]{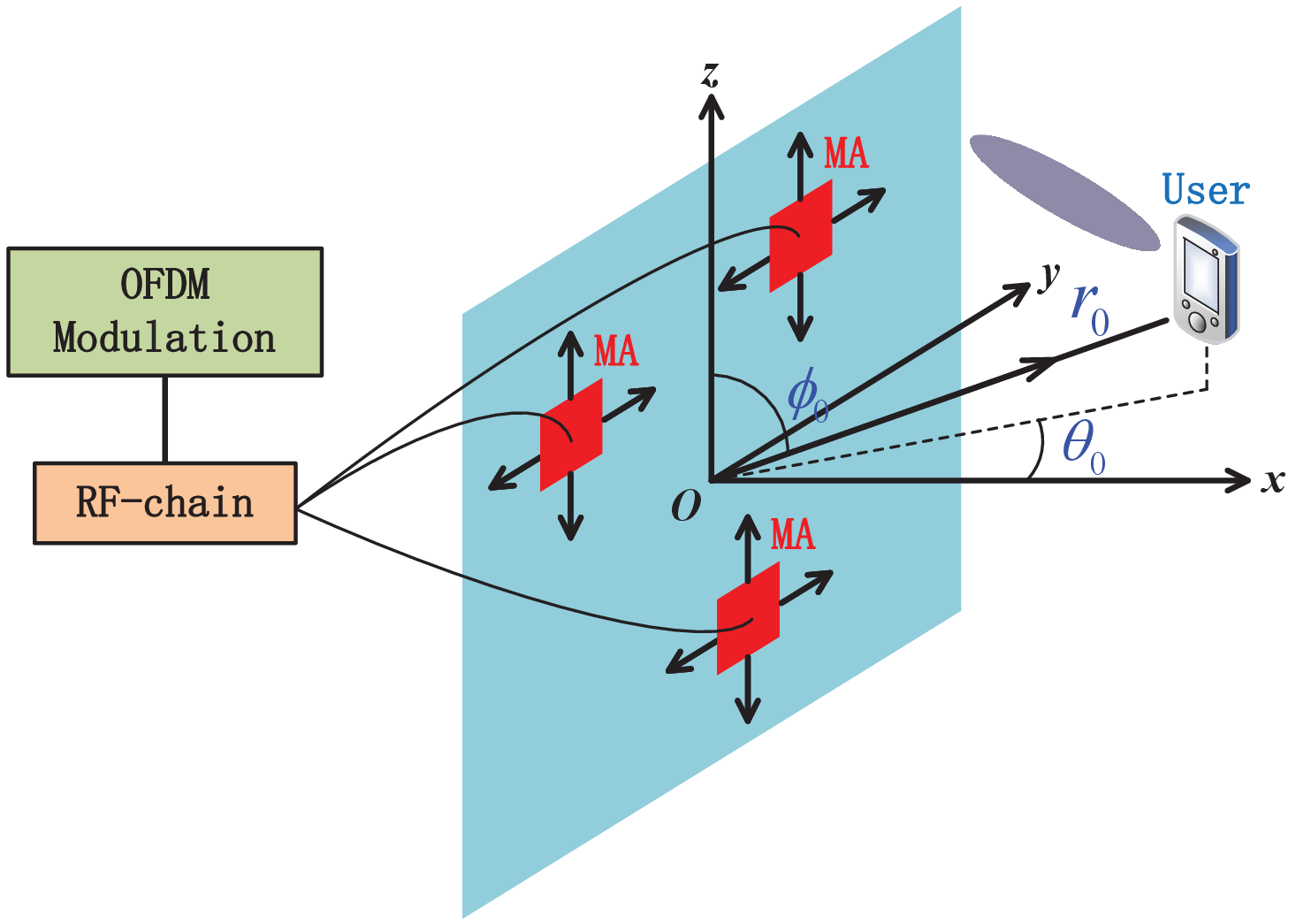}
\caption*{Fig. 1. Wideband MISO MA communication system.}
\end{figure}

In this paper, we consider quasi-static NF channel model. In fact, wideband systems generally operate in frequency band, e.g., mmWave and THz, and hence have large Rayleigh distance. This implies that wireless channels should be captured via more accurate NF channel model [10], [11]. Especially, in typical NF propagation environment, line-of-sight (LoS) path significantly dominates the non-LoS paths and therefore the latter are negligible [10], [11]. Therefore, we adopt NF LoS channel model in the following discussion. As specified in Fig. 1, regarding the origin as the reference point, we define $\theta_{0}$, $\phi_{0}$, and $r_{0}$ as the azimuth angle, elevation angle, and distance of the user's antenna, respectively.$^{1}$\footnote{$^{1}$Note $(\theta_{0}, \phi_{0}, r_{0})$ can be estimated by various existing sensing techniques, e.g., [13], [14]. Hence, we just assume that $(\theta_{0}, \phi_{0}, r_{0})$ is known in this paper.} Hence, the position of user's antenna is given by $\bm{p}_{\mathrm{U}} = [r_{0}\mathsf{cos}\theta_{0}\mathsf{sin}\phi_{0}, r_{0}\mathsf{sin}\theta_{0}\mathsf{sin}\phi_{0}, r_{0}\mathsf{cos}\phi_{0}]^{\mathsf{T}}$. Furthermore, given the MAs' positions $\{ \bm{p}_{m} \}_{m=1}^{M}$, the normalized NF channel response vector at frequency $f$ reads
\begin{align}
\bm{b}(\{ \bm{p}_{m} \}_{m=1}^{M}, \theta_{0}, \phi_{0}, r_{0}, f) = [&e^{-j\frac{2\pi f}{c}\| \bm{p}_{\mathrm{U}} - \bm{p}_{1} \|_{2}}, \notag\\
&\cdots, e^{-j\frac{2\pi f}{c}\| \bm{p}_{\mathrm{U}} - \bm{p}_{M} \|_{2}}]^{\mathsf{T}},
\end{align}
where $c$ is the speed of light. Based on Fresnel approximation [15], $\| \bm{p}_{\mathrm{U}} - \bm{p}_{m} \|_{2}, \; m \in \mathcal{M}$, can be appropriately approximated as$^{2}$\footnote{$^{2}$By setting $r_{0} \rightarrow \infty$, (1) reduces to conventional FF steering vector. Hence, all discussions in this paper can indeed be applied to FF scenario straightforwardly.}
\begin{align}
&\| \bm{p}_{\mathrm{U}} - \bm{p}_{m} \|_{2} \approx r_{0} - y_{m}\mathsf{sin}\theta_{0}\mathsf{sin}\phi_{0} - z_{m}\mathsf{cos}\phi_{0} \notag\\
& + \frac{y_{m}^{2} + z_{m}^{2} - (y_{m}\mathsf{sin}\theta_{0}\mathsf{sin}\phi_{0} + z_{m}\mathsf{cos}\phi_{0})^{2}}{2r_{0}}, \; m \in \mathcal{M}.
\end{align}

Analogous to conventional hybrid beamforming architecture, whose phase-shifter array determines the beam direction physically, MA system's analog beam is conquered by adjustable MAs' positions. To investigate the beam squint effect in wideband system, we consider the analog beamforming gain endowed by the MA array [8]-[11]. Specifically, denote the MA array's beam steering vector as $\bm{w}(\{ \bm{p}_{m} \}_{m=1}^{M}, \theta_{0}, \phi_{0}, r_{0}, f_{\mathrm{c}})$, which is similarly expressed in (1) with $f_{\mathrm{c}}$ being the central frequency. Then, the MAs' analog beamforming gain is
\begin{equation}
\mathsf{g}(\{ \bm{p}_{m} \}_{m\!=\!1}^{M}, f_{\mathrm{c}}, f) \!\!\triangleq\!\! |\bm{w}^{\mathsf{H}}(\{ \bm{p}_{m} \}_{m\!=\!1}^{M}, f_{\mathrm{c}})\bm{b}(\{ \bm{p}_{m} \}_{m\!=\!1}^{M}, f)|,
\end{equation}
where $\bm{w}(\{ \bm{p}_{m} \}_{m=1}^{M}, f_{\mathrm{c}})$ and $\bm{b}(\{ \bm{p}_{m} \}_{m=1}^{M}, f)$ respectively represent $\bm{w}(\{ \bm{p}_{m} \}_{m=1}^{M}, \theta_{0}, \phi_{0}, r_{0}, f_{\mathrm{c}})$ and $\bm{b}(\{ \bm{p}_{m} \}_{m=1}^{M}, \theta_{0}, \phi_{0}, r_{0}, f)$ for simplicity.

As indicated by (3), analog beamforming gain depends on the subcarrier frequency $f$. For the wideband system, the carrier frequency $f$ varies over a wide spectrum and hence incurs severe fluctuating beamforming gain across subcarriers, which is the so-called beam squint effect. Note that the squinted beam pattern is related to the path delay between each antenna and user [8]-[11], which can be affected by the positions of antennas. Hence, the beam squint effect could be alleviated via delicately configuring MAs' positions, in other words, $\mathsf{g}(\{ \bm{p}_{m} \}_{m=1}^{M}, f_{\mathrm{c}}, f)$ among different $f$'s can be improved. To provide more balanced communication rate among different subcarrier's data stream [8], [9], [16], we propose to maximize the minimum analog beamforming among the entire band. Specifically, denote the subcarrier frequencies as $\{ f_{l} \}_{l=0}^{L}$ with $f_{0} < \cdots < f_{L}$. The configuration of MAs is formulated as the following problem
\begin{align}
(\mathcal{P}1): \max_{\{ \bm{t}_{m} \}_{m=1}^{M}} &\min_{f \in \{ f_{l} \}_{l=0}^{L}} \mathsf{g}^{2}(\{ \bm{t}_{m} \}_{m=1}^{M}, f_{\mathrm{c}}, f) \\
\mathrm{s.t.} \; &[0, \bm{t}_{m}^{\mathsf{T}}]^{\mathsf{T}} \in \mathcal{C}_{\mathrm{MA}}, \; m \in \mathcal{M}, \tag{4a}\\
&\| \bm{t}_{m} - \bm{t}_{s} \|_{2} \geq D_{\mathrm{min}}, \; m, s \!\in\! \mathcal{M}, \; m \!\neq\! s, \tag{4b}
\end{align}
where $\bm{t}_{m} \triangleq [y_{m}, z_{m}]^{\mathsf{T}}, \; m \in \mathcal{M}$, constraint (4a) requires that all MAs should be in feasible region $\mathcal{C}_{\mathrm{MA}}$, and constraint (4b) ensures realistic spacing between MAs [4]-[7]. Obviously, $(\mathcal{P}1)$ is highly challenging due to its objective and nonconvex constraint (4b). In the following, we will develop algorithms to tackle $(\mathcal{P}1)$.

\section{SV-Based Classical Solution}

In this section, via introducing a slack variable (SV), we develop an algorithm to solve the max-min problem $(\mathcal{P}1)$. Specifically, $(\mathcal{P}1)$ can be equivalently transformed into
\begin{align}
(\mathcal{P}2): &\max_{\{ \bm{t}_{m} \}_{m=1}^{M}, \kappa} \kappa \\
\mathrm{s.t.} \; &\mathsf{g}^{2}(\{ \bm{t}_{m} \}_{m=1}^{M}, f_{\mathrm{c}}, f_{l}) \geq \kappa, \; l = 0, \cdots, L, \tag{5a}\\
&[0, \bm{t}_{m}^{\mathsf{T}}]^{\mathsf{T}} \in \mathcal{C}_{\mathrm{MA}}, \; m \in \mathcal{M}, \tag{5b}\\
&\| \bm{t}_{m} - \bm{t}_{s} \|_{2} \geq D_{\mathrm{min}}, \; m, s \in \mathcal{M}, \; m \neq s. \tag{5c}
\end{align}

To obtain a tractable solution, in the following, we adopt block coordinate descent (BCD) method to alternatively optimize one MA's position at a time. Considering the optimization of the $m$-th MA's position alone, $(\mathcal{P}2)$ reduces to
\begin{align}
(\mathcal{P}3): \max_{\bm{t}_{m}, \kappa} \; &\kappa \\
\mathrm{s.t.} \; &\mathsf{g}^{2}(\bm{t}_{m}, f_{\mathrm{c}}, f_{l}) \geq \kappa, \; l = 0, \cdots, L, \tag{6a}\\
&[0, \bm{t}_{m}^{\mathsf{T}}]^{\mathsf{T}} \in \mathcal{C}_{\mathrm{MA}}, \tag{6b}\\
&\| \bm{t}_{m} - \bm{t}_{s} \|_{2} \geq D_{\mathrm{min}}, \; s \in \mathcal{M}, \; s \neq m. \tag{6c}
\end{align}

Note $(\mathcal{P}3)$ is still difficult due to the nonconvex constraints (6a) and (6c). To tackle this issue, we leverage majorization-minimization (MM) methodology [17] to construct concave lower bounds for the left hand side of (6a) and (6c), respectively. Note in the following, we will use $\mathsf{h}(\bm{t}_{m}, f_{l}) \triangleq \mathsf{g}^{2}(\bm{t}_{m}, f_{\mathrm{c}}, f_{l}), \; m \in \mathcal{M}, \; l = 0, \cdots, L$, to simplify notations.

Firstly, to handle (6a), we examine the second-order Taylor expansion of the left hand side of (6a) and replace its Hessian matrix with a smaller one in the sense of Loewner ordering to obtain a tight globally lower-bound surrogate. Specifically, the second-order Taylor expansion of $\mathsf{h}(\bm{t}_{m}, f_{l})$ with respect to (w.r.t.) $\bm{t}_{m}$ at the point $\bm{t}_{m}^{(n)}$ reads
\begin{align}
&\mathsf{h}(\bm{t}_{m}, f_{l}) \approx \mathsf{h}(\bm{t}_{m}^{(n)}, f_{l}) + \nabla_{\bm{t}_{m}}^{\mathsf{T}}(\mathsf{h}(\bm{t}_{m}, f_{l}))|_{\bm{t}_{m} = \bm{t}_{m}^{(n)}}(\bm{t}_{m} - \bm{t}_{m}^{(n)}) \notag\\
& + \frac{1}{2}(\bm{t}_{m} - \bm{t}_{m}^{(n)})^{\mathsf{T}}\nabla_{\bm{t}_{m}}^{2}(\mathsf{h}(\bm{t}_{m}, f_{l}))|_{\bm{t}_{m} = \bm{t}_{m}^{(n)}}(\bm{t}_{m} - \bm{t}_{m}^{(n)}),
\end{align}
where $\nabla_{\bm{t}_{m}}(\mathsf{h}(\bm{t}_{m}, f_{l}))$ and $\nabla_{\bm{t}_{m}}^{2}(\mathsf{h}(\bm{t}_{m}, f_{l}))$ are derived in Appendix A. Furthermore, it can be observed that $\mathsf{h}(\bm{t}_{m}, f_{l})$ has bounded curvature, i.e., there must exist a negative semidefinite matrix $\mathbf{M}_{m,l}$ such that $\mathbf{M}_{m,l} \preceq \nabla_{\bm{t}_{m}}^{2}(\mathsf{h}(\bm{t}_{m}, f_{l}))$. Therefore, a global concave lower bound of $\mathsf{h}(\bm{t}_{m}, f_{l})$ can be obtained by substituting $\mathbf{M}_{m,l}$ into (7). Detailed derivation of $\mathbf{M}_{m,l}$ is given in Appendix B.

Next, we convexify the constraint (6c), which is much easier since the left hand side of (6c) is convex w.r.t. $\bm{t}_{m}$. By directly utilizing the following first-order Taylor expansion
\begin{equation}
\| \bm{t}_{m} - \bm{t}_{s} \|_{2} \geq \frac{(\bm{t}_{m}^{(n)} - \bm{t}_{s})^{\mathsf{T}}}{\| \bm{t}_{m}^{(n)} - \bm{t}_{s} \|_{2}}(\bm{t}_{m} - \bm{t}_{s}), \; s \in \mathcal{M}, \; s \neq m,
\end{equation}
a concave lower bound of $\| \bm{t}_{m} - \bm{t}_{s} \|_{2}$ can be obtained.

After surrogating the left hand side of (6a) and (6c) by the two lower bounds developed above, we acquire the following convex optimization problem
\begin{align}
(\mathcal{P}4): &\max_{\bm{t}_{m}, \kappa} \; \kappa \\
\mathrm{s.t.} \; &\mathsf{h}(\bm{t}_{m}, f_{l}|\bm{t}_{m}^{(n)}) \geq \kappa, \; l = 0, \cdots, L, \tag{9a}\\
&[0, \bm{t}_{m}^{\mathsf{T}}]^{\mathsf{T}} \in \mathcal{C}_{\mathrm{MA}}, \tag{9b}\\
&\frac{(\bm{t}_{m}^{(n)} - \bm{t}_{s})^{\mathsf{T}}}{\| \bm{t}_{m}^{(n)} - \bm{t}_{s} \|_{2}}(\bm{t}_{m} - \bm{t}_{s}) \geq D_{\mathrm{min}}, \; s \in \mathcal{M}, \; s \neq m, \tag{9c}
\end{align}
which can be solved by numerical solvers, e.g., CVX, where
\begin{align}
\mathsf{h}(\bm{t}_{m}, f_{l}|\bm{t}_{m}^{(n)}) &\!\!\triangleq\!\! \mathsf{h}(\bm{t}_{m}^{(n)}, f_{l}) \!\!+\!\! \nabla_{\bm{t}_{m}}^{\mathsf{T}}(\mathsf{h}(\bm{t}_{m}, f_{l}))|_{\bm{t}_{m} = \bm{t}_{m}^{(n)}}(\bm{t}_{m} \!\!-\!\! \bm{t}_{m}^{(n)}) \notag\\
& \!\!+\!\! \frac{1}{2}(\bm{t}_{m} - \bm{t}_{m}^{(n)})^{\mathsf{T}}\mathbf{M}_{m,l}(\bm{t}_{m} - \bm{t}_{m}^{(n)}).
\end{align}

The proposed solution to $(\mathcal{P}1)$ by introducing an SV is summarized in Algorithm 1.

\begin{algorithm}[!t]
\caption{Solving $(\mathcal{P}1)$ based on SV}
\begin{algorithmic}[1]
\STATE Initialize feasible $\{ \bm{t}_{m}^{(0)} \}_{m=1}^{M}$ and $n = 0$;
\REPEAT
\FOR{$m = 1$ to $M$}
\STATE update $(\bm{t}_{m}^{(n + 1)}, \kappa^{(n + \frac{m}{M})})$ by solving $(\mathcal{P}4)$;
\ENDFOR
\STATE $n := n + 1$;
\UNTIL{convergence}
\end{algorithmic}
\end{algorithm}

\section{SGDA-Based Efficient Solution}

Note one potential defect of Alg. 1 proposed above lies in its complexity. In fact, wideband system generally has a great number, e.g., up to hundreds or thousands, of subcarriers. This makes $(\mathcal{P}4)$ a second-order cone programming (SOCP) problem with a large number of constraints and hence significantly inflates its computational complexity. In this section, we proceed to develop an efficient algorithm based on SGDA method.

We still exploit BCD framework to optimize the antenna positions. Before invoking SGDA algorithm, we convexify constraint (4b) by (8), and regard $f$ as a continuous parameter, i.e., $f \in [f_{0}, f_{L}]$.$^{3}$\footnote{$^{3}$This is especially reasonable for wideband system since the entire spectrum is finely notched in this case.} Hence, the optimization for the $m$-th MA's position can be casted as
\begin{align}
(\mathcal{P}5): &\max_{\bm{t}_{m}} \min_{f \in [f_{0}, f_{L}]} \mathsf{h}(\bm{t}_{m}, f) \\
\mathrm{s.t.} \; &[0, \bm{t}_{m}^{\mathsf{T}}]^{\mathsf{T}} \in \mathcal{C}_{\mathrm{MA}}, \tag{11a}\\
&\frac{(\bm{t}_{m}^{(n)} \!-\! \bm{t}_{s})^{\mathsf{T}}}{\| \bm{t}_{m}^{(n)} \!-\! \bm{t}_{s} \|_{2}}(\bm{t}_{m} \!-\! \bm{t}_{s}) \geq D_{\mathrm{min}}, \; s \in \mathcal{M}, \; s \neq m. \tag{11b}
\end{align}

Next, we adopt SGDA method to tackle $(\mathcal{P}5)$. Note GDA methodology solves max-min problem via alternatively conducting gradient ascent (GA) and gradient descent (GD) updates towards maximization and minimization, respectively. Besides, by introducing additional smoothing terms, SGDA further improves the convergence of the standard GDA method [12]. Specifically, according to SGDA method [12], by introducing auxiliary variables $\{ \bm{\alpha}_{m} \}_{m=1}^{M}$ and defining a function
\begin{equation}
\mathsf{k}(\bm{t}_{m}, \bm{\alpha}_{m};f) \triangleq \mathsf{h}(\bm{t}_{m}, f) + \frac{p_{\mathrm{S}}}{2}\| \bm{t}_{m} - \bm{\alpha}_{m} \|_{2}^{2},
\end{equation}
where $p_{\mathrm{S}} > 0$ is a pre-defined constant, we alternatively update $\bm{t}_{m}$ in a GA manner while updating $f$ and $\bm{\alpha}_{m}$ in GD manners in each SGDA iteration. Denote $(\bm{t}_{m}^{(n,i)}, f^{(n,i)}, \bm{\alpha}_{m}^{(n,i)})$ as the solution iterate updated in the $i$-th SGDA iteration during the $n$-th BCD iteration. The detailed updates will be elaborated in the following discussions.

\subsection*{A. Update of $\bm{t}_{m}$}

The variable $\bm{t}_{m}$ is updated in a GA manner, given as follows
\begin{equation}
\bm{t}_{m}^{(n,i + 1)} \!\!:=\!\! \bm{t}_{m}^{(n,i)} \!+\! \eta_{\bm{t}_{m}}^{(n,i)}\nabla_{\bm{t}_{m}}\!(\mathsf{k}(\bm{t}_{m}, \bm{\alpha}_{m}^{(n,i)};f^{(n,i)}\!)\!)\!|_{\bm{t}_{m} \!= \bm{t}_{m}^{(n,i)}},
\end{equation}
where $\eta_{\bm{t}_{m}}^{(n,i)}$ is the updating step size and can be set as a constant [12], and $\nabla_{\bm{t}_{m}}(\mathsf{k}(\bm{t}_{m}, \bm{\alpha}_{m};f))$ is
\begin{equation}
\nabla_{\bm{t}_{m}}(\mathsf{k}(\bm{t}_{m}, \bm{\alpha}_{m};f)) = \nabla_{\bm{t}_{m}}(\mathsf{h}(\bm{t}_{m}, f)) + p_{\mathrm{S}}(\bm{t}_{m} - \bm{\alpha}_{m}),
\end{equation}
where $\nabla_{\bm{t}_{m}}(\mathsf{h}(\bm{t}_{m}, f))$ has been derived in Appendix A.

Note that the newly obtained $\bm{t}_{m}^{(n,i + 1)}$ may not satisfy (11a) or (11b). If so, $\bm{t}_{m}^{(n,i + 1)}$ must be projected onto the convex domain constituted by (11a) and (11b), which is meant to solve the following optimization
\begin{align}
(\mathcal{P}6): &\min_{\bm{x} \in \mathbb{R}^{2}} \; \frac{1}{2}\| \bm{x} - \bm{t}_{m}^{(n,i + 1)}\|_{2}^{2} \\
\mathrm{s.t.} \; &[0, \bm{x}^{\mathsf{T}}]^{\mathsf{T}} \in \mathcal{C}_{\mathrm{MA}}, \tag{15a}\\
&\frac{(\bm{t}_{m}^{(n)} \!-\! \bm{t}_{s})^{\mathsf{T}}}{\| \bm{t}_{m}^{(n)} \!-\! \bm{t}_{s} \|_{2}}(\bm{x} \!-\! \bm{t}_{s}) \geq D_{\mathrm{min}}, \; s \in \mathcal{M}, \; s \neq m. \tag{15b}
\end{align}
Problem $(\mathcal{P}6)$ is convex which can be numerically solved.

\subsection*{B. Update of $f$}

The variable $f$ is updated in a GD manner, i.e.,
\begin{equation}
f^{(n,i + 1)} \!\!:=\!\! f^{(n,i)} \!\!-\!\! \eta_{f}^{(n,i)}\nabla_{f}(\mathsf{k}(\bm{t}_{m}^{(n,i + 1)}, \bm{\alpha}_{m}^{(n,i)};f)\!)\!|_{f \!=\! f^{(n,i)}},
\end{equation}
where $\eta_{f}^{(n,i)}$ is the updating step size and can also be set as a constant [12]. The calculation of $\nabla_{f}(\mathsf{k}(\bm{t}_{m}, \bm{\alpha}_{m};f))$ is relegated to Appendix C.

Obviously, there exists a possibility that $f^{(n,i + 1)}$ falls out of the feasible domain $[f_{0}, f_{L}]$. When this occurs, we should project $f^{(n,i + 1)}$ onto $[f_{0}, f_{L}]$, that is:
\begin{itemize}

\item If $f^{(n,i + 1)} < f_{0}$, $f^{(n,i + 1)} = f_{0}$,

\item or if $f^{(n,i + 1)} > f_{L}$, $f^{(n,i + 1)} = f_{L}$.

\end{itemize}

\subsection*{C. Update of $\bm{\alpha}_{m}$}

The variable $\bm{\alpha}_{m}$ is updated as follows [12]
\begin{equation}
\bm{\alpha}_{m}^{(n,i + 1)} := \bm{\alpha}_{m}^{(n,i)} + \eta_{\bm{\alpha}_{m}}^{(n,i)}(\bm{t}_{m}^{(n,i + 1)} - \bm{\alpha}_{m}^{(n,i)}),
\end{equation}
where $\eta_{\bm{\alpha}_{m}}^{(n,i)} \in (0, 1]$ is a constant.

The proposed SGDA-based algorithm to tackle $(\mathcal{P}1)$ is summarized in Algorithm 2.

\begin{algorithm}[!t]
\caption{Solving $(\mathcal{P}1)$ based on SGDA}
\begin{algorithmic}[1]
\STATE Initialize feasible $\{ \bm{t}_{m}^{(0)} \}_{m=1}^{M}$ and $n = 0$;
\REPEAT
\FOR{$m = 1$ to $M$}
\STATE set $\bm{t}_{m}^{(n,0)} := \bm{t}_{m}^{(n)}$, $\bm{\alpha}_{m}^{(n,0)} := \bm{t}_{m}^{(n,0)}$ and $i = 0$;
\STATE randomly generate $f^{(n,0)} \in [f_{0}, f_{L}]$;
\REPEAT
\STATE update $\bm{t}_{m}^{(n,i + 1)}$ by (13);
\STATE project $\bm{t}_{m}^{(n,i + 1)}$ by solving $(\mathcal{P}6)$ (if needed);
\STATE update $f^{(n,i + 1)}$ by (16);
\STATE project $f^{(n,i + 1)}$ by $f^{(n,i + 1)} = f_{0}$ \\ or $f^{(n,i + 1)} = f_{L}$ (if needed);
\STATE update $\bm{\alpha}_{m}^{(n,i + 1)}$ by (17);
\STATE $i := i + 1$;
\UNTIL{convergence}
\STATE set $\bm{t}_{m}^{(n + 1)} := \bm{t}_{m}^{(n,\infty)}$;
\ENDFOR
\STATE $n := n + 1$;
\UNTIL{convergence}
\end{algorithmic}
\end{algorithm}

\section{Simulation Results}

In this section, numerical results are provided to demonstrate the effectiveness of our proposed algorithms and the benefit of deploying MA array in wideband communication system. Unless otherwise specified, $L = 256$, $f_{0} = 58.92\mathrm{GHz}$, $f_{L} = 61.08\mathrm{GHz}$, $f_{\mathrm{c}} = \frac{f_{0} + f_{L}}{2} = 60\mathrm{GHz}$ [18], $f_{l} = f_{0} + \frac{l}{L}(f_{L} - f_{0}), \; l = 0, \cdots, L$, $A = \frac{100c}{f_{c}}$ [19], $D_{\mathrm{min}} = \frac{c}{2f_{0}}$, $\theta_{0} = \frac{\pi}{6}$, $\phi_{0} = \frac{\pi}{4}$, $r_{0} = 10\mathrm{m}$, $p_{\mathrm{S}} = 2$, $\eta_{\bm{\alpha}_{m}} = 0.5, \; m \in \mathcal{M}$.

\begin{figure}[!t]
\centering
\includegraphics[scale=0.32]{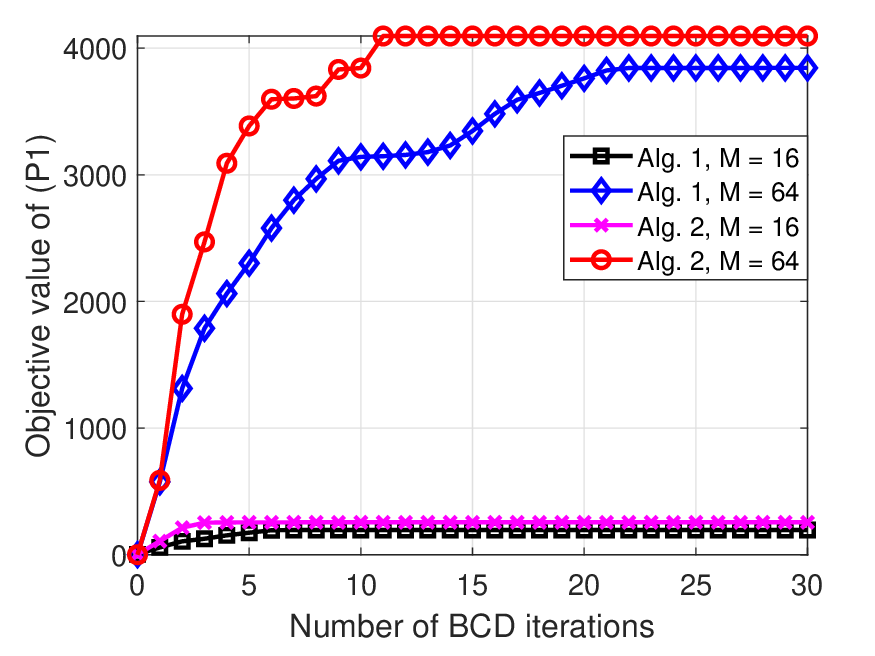}
\caption*{Fig. 2. Convergence behaviours of Alg. 1 and Alg. 2.}
\end{figure}

\begin{figure}[!t]
\centering
\begin{subfigure}[c]{1\linewidth}
\centering
\includegraphics[scale=0.32]{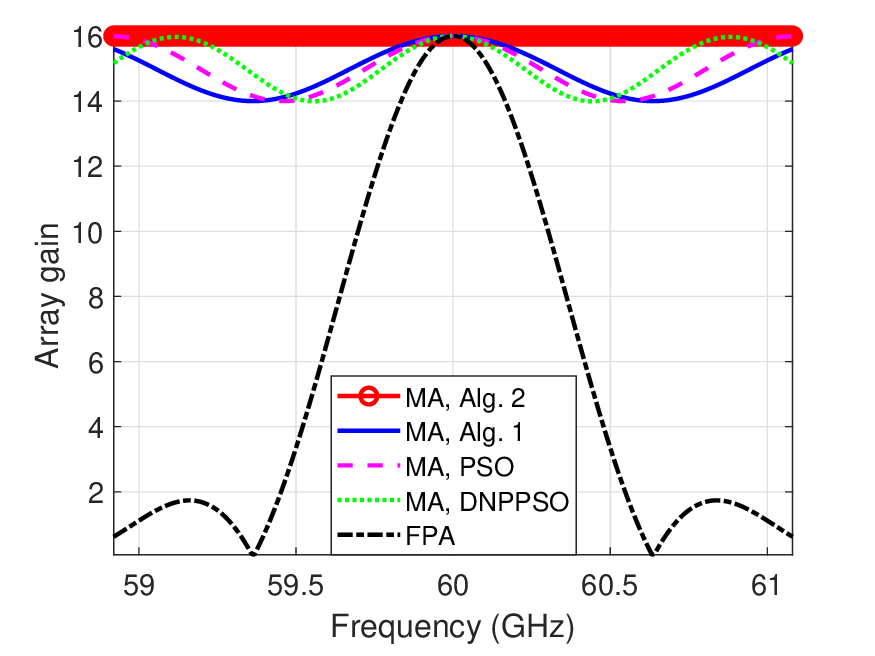}
\caption*{(a) $M = 16$.}
\end{subfigure}
\begin{subfigure}[c]{1\linewidth}
\centering
\includegraphics[scale=0.32]{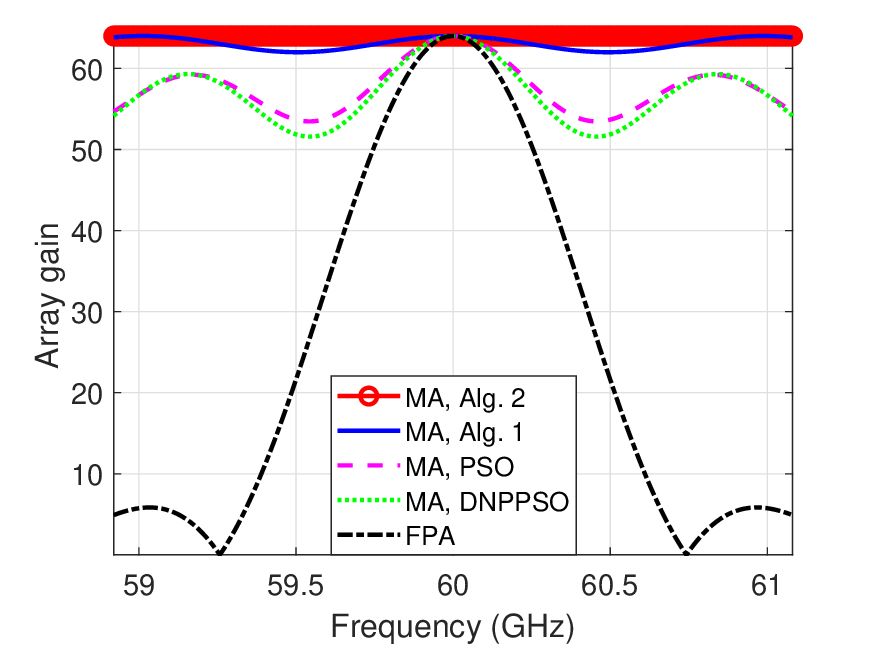}
\caption*{(b) $M = 64$.}
\end{subfigure}
\caption*{Fig. 3. Array gain across considered frequency band.}
\end{figure}

\begin{figure}[!t]
\centering
\includegraphics[scale=0.32]{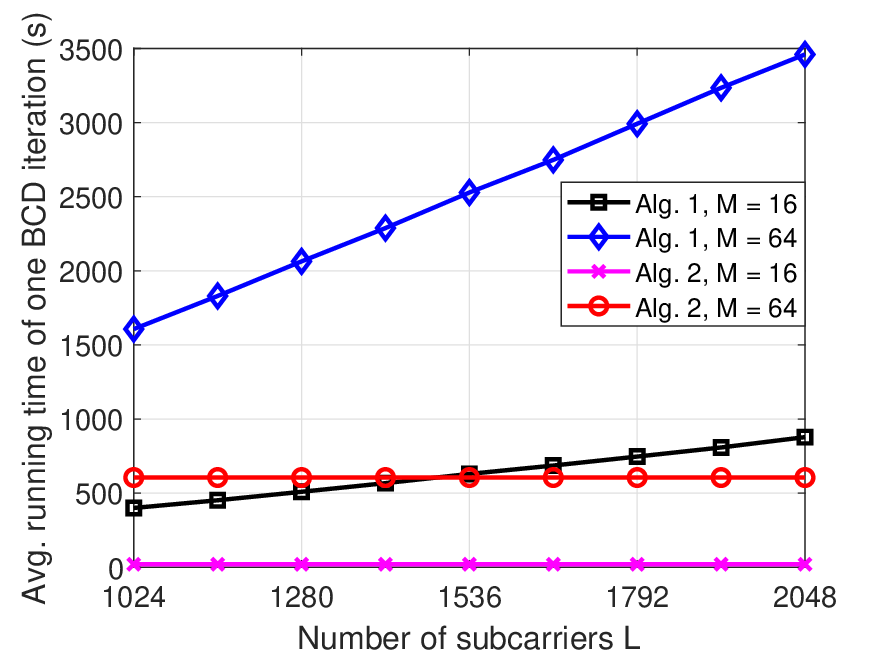}
\caption*{Fig. 4. Comparison of run-time between Alg. 1 and Alg. 2.}
\end{figure}

Fig. 2 illustrates the convergence behaviours of Alg. 1 and Alg. 2. As suggested by Fig. 2, by utilizing any one of our proposed solutions, the objective value of $(\mathcal{P}1)$ monotonically increases and converges in tens of iterations. Interestingly, besides its lower complexity (see Fig. 4), Alg. 2 yields superior converged objective values over Alg. 1.

Fig. 3 plots the array gain (defined in (3)) across different subcarriers. Besides our proposed methods, standard particle swarm optimization (PSO) and advanced dynamic neighborhood pruning PSO (DNPPSO) [19] frameworks for solving $(\mathcal{P}1)$ are also exploited as benchmarks. As reflected by Fig. 3, via appropriately configuring the antennas' positions, MA can remarkably mitigate the beam squint effect compared to the conventional FPA counterpart. Note that the analog beamforming gain yielded by Alg. 2 outperforms Alg. 1 over the entire wide frequency band. In addition, our proposed schemes exhibit superior performances over PSO-based counterparts.

Lastly, to examine the complexity of our proposed algorithms, we compare both average MATLAB run-time per BCD iteration in Fig. 4. As reflected by Fig. 4, the run-time of Alg. 1 increases intensively when the number of subcarriers $L$ grows. In contrast, the run-time per BCD iteration of Alg. 2 is nearly insensitive to $L$ and maintains a much lower level than that of Alg. 1.

\section{Conclusion}

In this paper, a wideband MISO communication system aided by MA array under NF channel model is considered. To overcome beam squint effect, a minimal array gain maximization criterion is proposed to configure MAs' positions. We successfully develop algorithms based on introducing an SV and SGDA method. The SGDA-based algorithm has much lower complexity and is especially suitable to systems with large bandwidth. Simulation results verify that beam squint effect can be significantly mitigated by MA array deployment.

Although this paper only considers a single-user scenario, the proposed scheme for tackling beam squint issue can be extended to multi-user systems, where other objectives, e.g., spectral efficiency, rather than array gain, may be considered due to multi-user interference. Besides, the latest proposed six-dimensional movable antenna (6DMA), which offers both movability and rotatability, could further enhance beam squint mitigation [20]. These open problems are worthy to be investigated in future researches.

\appendix

\subsection*{A. Calculation of $\nabla_{\bm{t}_{m}}(\mathsf{h}(\bm{t}_{m}, f_{l}))$ and $\nabla_{\bm{t}_{m}}^{2}(\mathsf{h}(\bm{t}_{m}, f_{l}))$}

Define $C_{m,l} \triangleq \sum_{n=1, n \neq m}^{M}e^{jF_{l}\mathsf{a}(\bm{t}_{n})}, \; m \in \mathcal{M}, \; l = 0, \cdots, L$, where $F_{l} \triangleq \frac{2\pi(f_{\mathrm{c}} - f_{l})}{c}, \; l = 0, \cdots, L$, $\mathsf{a}(\bm{t}_{m}) \triangleq r_{0} - y_{m}w_{0} - z_{m}u_{0} + \frac{y_{m}^{2} + z_{m}^{2} - (y_{m}w_{0} + z_{m}u_{0})^{2}}{2r_{0}}, \; m \in \mathcal{M}$, $w_{0} \triangleq \mathsf{sin}\theta_{0}\mathsf{sin}\phi_{0}$ and $u_{0} \triangleq \mathsf{cos}\phi_{0}$. Then, after some manipulations, $\mathsf{h}(\bm{t}_{m}, f_{l})$ can be expressed as
\begin{equation}
\mathsf{h}(\bm{t}_{m}, f_{l}) \!=\! 2|C_{m,l}|\mathsf{cos}(F_{l}\mathsf{a}(\bm{t}_{m}) \!-\! \angle(C_{m,l})) \!+\! |C_{m,l}|^{2} \!+\! 1,
\end{equation}
where $\angle(x)$ picks up the phase of input complex scalar $x$.

The gradient $\nabla_{\bm{t}_{m}}(\mathsf{h}(\bm{t}_{m}, f_{l}))$ is given by
\begin{equation}
\nabla_{\bm{t}_{m}}(\mathsf{h}(\bm{t}_{m}, f_{l})) = \bigg[\frac{\partial\mathsf{h}(\bm{t}_{m}, f_{l})}{\partial y_{m}}, \frac{\partial\mathsf{h}(\bm{t}_{m}, f_{l})}{\partial z_{m}}\bigg]^{\mathsf{T}},
\end{equation}
where
\begin{align}
&\frac{\partial\mathsf{h}(\bm{t}_{m}, f_{l})}{\partial y_{m}} \!\!=\!\! -2|C_{m,l}|F_{l}\frac{\partial\mathsf{a}(\bm{t}_{m})}{\partial y_{m}}\mathsf{sin}(F_{l}\mathsf{a}(\bm{t}_{m}) \!\!-\!\! \angle(C_{m,l})), \\
&\frac{\partial\mathsf{a}(\bm{t}_{m})}{\partial y_{m}} = \frac{y_{m} - (y_{m}w_{0} + z_{m}u_{0})w_{0}}{r_{0}} - w_{0}, \\
&\frac{\partial\mathsf{h}(\bm{t}_{m}, f_{l})}{\partial z_{m}} \!\!=\!\! -2|C_{m,l}|F_{l}\frac{\partial\mathsf{a}(\bm{t}_{m})}{\partial z_{m}}\mathsf{sin}(F_{l}\mathsf{a}(\bm{t}_{m}) \!\!-\!\! \angle(C_{m,l})), \\
&\frac{\partial\mathsf{a}(\bm{t}_{m})}{\partial z_{m}} = \frac{z_{m} - (y_{m}w_{0} + z_{m}u_{0})u_{0}}{r_{0}} - u_{0}.
\end{align}

The Hessian matrix $\nabla_{\bm{t}_{m}}^{2}(\mathsf{h}(\bm{t}_{m}, f_{l}))$ is given as follows
\begin{equation}
\nabla_{\bm{t}_{m}}^{2}(\mathsf{h}(\bm{t}_{m}, f_{l})) = \begin{bmatrix}
\frac{\partial^{2}\mathsf{h}(\bm{t}_{m}, f_{l})}{\partial y_{m}^{2}} & \frac{\partial^{2}\mathsf{h}(\bm{t}_{m}, f_{l})}{\partial y_{m}\partial z_{m}} \\
\frac{\partial^{2}\mathsf{h}(\bm{t}_{m}, f_{l})}{\partial z_{m}\partial y_{m}} & \frac{\partial^{2}\mathsf{h}(\bm{t}_{m}, f_{l})}{\partial z_{m}^{2}} \\
\end{bmatrix},
\end{equation}
where
\begin{align}
&\frac{\partial^{2}\mathsf{h}(\bm{t}_{m}, f_{l})}{\partial y_{m}^{2}} = -2|C_{m,l}|F_{l}\frac{1 - w_{0}^{2}}{r_{0}}\mathsf{sin}(F_{l}\mathsf{a}(\bm{t}_{m}) - \angle(C_{m,l})) \notag\\
& - 2|C_{m,l}|F_{l}^{2}\bigg(\frac{\partial\mathsf{a}(\bm{t}_{m})}{\partial y_{m}}\bigg)^{2}\mathsf{cos}(F_{l}\mathsf{a}(\bm{t}_{m}) - \angle(C_{m,l})), \\
&\frac{\partial^{2}\mathsf{h}(\bm{t}_{m}, f_{l})}{\partial y_{m}\partial z_{m}} = -2|C_{m,l}|F_{l}\frac{-w_{0}u_{0}}{r_{0}}\mathsf{sin}(F_{l}\mathsf{a}(\bm{t}_{m}) - \angle(C_{m,l})) \notag\\
& - 2|C_{m,l}|F_{l}^{2}\frac{\partial\mathsf{a}(\bm{t}_{m})}{\partial y_{m}}\frac{\partial\mathsf{a}(\bm{t}_{m})}{\partial z_{m}}\mathsf{cos}(F_{l}\mathsf{a}(\bm{t}_{m}) - \angle(C_{m,l})), \\
&\frac{\partial^{2}\mathsf{h}(\bm{t}_{m}, f_{l})}{\partial z_{m}\partial y_{m}} = \frac{\partial^{2}\mathsf{h}(\bm{t}_{m}, f_{l})}{\partial y_{m}\partial z_{m}}, \\
&\frac{\partial^{2}\mathsf{h}(\bm{t}_{m}, f_{l})}{\partial z_{m}^{2}} = -2|C_{m,l}|F_{l}\frac{1 - u_{0}^{2}}{r_{0}}\mathsf{sin}(F_{l}\mathsf{a}(\bm{t}_{m}) - \angle(C_{m,l})) \notag\\
& - 2|C_{m,l}|F_{l}^{2}\bigg(\frac{\partial\mathsf{a}(\bm{t}_{m})}{\partial z_{m}}\bigg)^{2}\mathsf{cos}(F_{l}\mathsf{a}(\bm{t}_{m}) - \angle(C_{m,l})).
\end{align}

\subsection*{B. Derivation of a Lower Bound $\mathbf{M}_{m,l}$ for Hessian in (7)}

Denote $H_{1,m,l} \triangleq \frac{\partial^{2}\mathsf{h}(\bm{t}_{m}, f_{l})}{\partial y_{m}^{2}}$, $H_{2,m,l} \triangleq \frac{\partial^{2}\mathsf{h}(\bm{t}_{m}, f_{l})}{\partial y_{m}\partial z_{m}}$ and $H_{3,m,l} \triangleq \frac{\partial^{2}\mathsf{h}(\bm{t}_{m}, f_{l})}{\partial z_{m}^{2}}$. Then, the Hessian matrix $\nabla_{\bm{t}_{m}}^{2}(\mathsf{h}(\bm{t}_{m}, f_{l}))$ can be recasted as
\begin{equation}
\nabla_{\bm{t}_{m}}^{2}\!(\mathsf{h}(\bm{t}_{m}, f_{l})) \!\!=\!\! \underbrace{\begin{bmatrix}
H_{1,m,l} & 0 \\
0 & H_{3,m,l} \\
\end{bmatrix}}_{\mathbf{H}_{1,m,l}} \!\!+\!\! \underbrace{\begin{bmatrix}
0 & H_{2,m,l} \\
H_{2,m,l} & 0 \\
\end{bmatrix}}_{\mathbf{H}_{2,m,l}}.
\end{equation}
Hence, the task of calculating $\mathbf{M}_{m,l}$ reduces to exploring the lower bounds of $\mathbf{H}_{1,m,l}$ and $\mathbf{H}_{2,m,l}$, respectively.

Taking $\mathbf{H}_{1,m,l}$ into account first, the lower bound of $\mathbf{H}_{1,m,l}$ can be obtained by examining the lower bounds of $H_{1,m,l}$ and $H_{3,m,l}$. For $H_{1,m,l}$, since $\mathsf{sin}(\cdot) \in [-1, 1]$ and $\mathsf{cos}(\cdot) \in [-1, 1]$, we have
\begin{equation}
H_{1,m,l} \!\geq\! -\bigg|2|C_{m,l}|F_{l}\frac{1 \!-\! w_{0}^{2}}{r_{0}}\bigg| \!-\! 2|C_{m,l}|F_{l}^{2}\bigg(\frac{\partial\mathsf{a}(\bm{t}_{m})}{\partial y_{m}}\bigg)^{2}.
\end{equation}
Note the right hand side of (30) still depends on $\bm{t}_{m}$, hence, the upper bound of $\big(\frac{\partial\mathsf{a}(\bm{t}_{m})}{\partial y_{m}}\big)^{2}$ needs to be evaluated. Due to the MAs' moving region constraint (6b), it can be verified that
\begin{equation}
\frac{\partial\mathsf{a}(\bm{t}_{m})}{\partial y_{m}} \!\!\in\!\! \bigg[\!-\!\frac{A}{2}\frac{1 \!\!-\!\! w_{0}^{2} \!\!+\!\! |w_{0}u_{0}|}{r_{0}} \!-\! w_{0}, \frac{A}{2}\frac{1 \!\!-\!\! w_{0}^{2} \!\!+\!\! |w_{0}u_{0}|}{r_{0}} \!-\! w_{0}\!\bigg].
\end{equation}
Therefore, it can be easily shown (with details omitted) that $\big(\frac{\partial\mathsf{a}(\bm{t}_{m})}{\partial y_{m}}\big)^{2}$ can be upper-bounded as follows
\begin{equation}
\bigg(\frac{\partial\mathsf{a}(\bm{t}_{m})}{\partial y_{m}}\bigg)^{2} \leq \bigg(\frac{A}{2}\frac{1 - w_{0}^{2} + |w_{0}u_{0}|}{r_{0}} + |w_{0}|\bigg)^{2}.
\end{equation}
Finally, via replacing $\big(\frac{\partial\mathsf{a}(\bm{t}_{m})}{\partial y_{m}}\big)^{2}$ in the right hand side of (30) by (32), we obtain a lower bound of $H_{1,m,l}$, given by
\begin{align}
H_{1,m,l} &\geq -\bigg|2|C_{m,l}|F_{l}\frac{1 - w_{0}^{2}}{r_{0}}\bigg| \notag\\
& - 2|C_{m,l}|F_{l}^{2}\bigg(\frac{A}{2}\frac{1 - w_{0}^{2} + |w_{0}u_{0}|}{r_{0}} + |w_{0}|\bigg)^{2},
\end{align}
which is indeed a global lower bound because the right hand of (33) is unrelated to $\bm{t}_{m}$.

Following similar arguments, a global lower bound of $H_{3,m,l}$ can be constructed as
\begin{align}
H_{3,m,l} &\geq -\bigg|2|C_{m,l}|F_{l}\frac{1 - u_{0}^{2}}{r_{0}}\bigg| \notag\\
& - 2|C_{m,l}|F_{l}^{2}\bigg(\frac{A}{2}\frac{1 - u_{0}^{2} + |w_{0}u_{0}|}{r_{0}} + |u_{0}|\bigg)^{2},
\end{align}
ans thus a global lower bound of $\mathbf{H}_{1,m,l}$ has been acquired.

Next, we focus on $\mathbf{H}_{2,m,l}$. Note that $\mathbf{H}_{2,m,l}$ naturally has the following lower bound
\begin{equation}
\mathbf{H}_{2,m,l} \succeq -|H_{2,m,l}|\mathbf{I}_{2},
\end{equation}
where $\mathbf{I}_{2}$ represents the identity matrix of size $2 \times 2$. Therefore, the global lower bound of $\mathbf{H}_{2,m,l}$ can be obtained by checking the upper bound of $|H_{2,m,l}|$. Since $\mathsf{sin}(\cdot) \in [-1, 1]$ and $\mathsf{cos}(\cdot) \in [-1, 1]$, $|H_{2,m,l}|$ can be upper-bounded by
\begin{align}
&|H_{2,m,l}| \leq \bigg|2|C_{m,l}|F_{l}\frac{w_{0}u_{0}}{r_{0}}\bigg| + 2|C_{m,l}|F_{l}^{2} \notag\\
& \times \mathsf{max} \bigg\{ \bigg|\bigg[\frac{\partial\mathsf{a}(\bm{t}_{m})}{\partial y_{m}}\frac{\partial\mathsf{a}(\bm{t}_{m})}{\partial z_{m}}\bigg]_{\!\mathrm{L}}\bigg|, \bigg|\bigg[\frac{\partial\mathsf{a}(\bm{t}_{m})}{\partial y_{m}}\frac{\partial\mathsf{a}(\bm{t}_{m})}{\partial z_{m}}\bigg]_{\!\mathrm{U}}\bigg| \bigg\},
\end{align}
where $[x]_{\mathrm{L}}$ and $[x]_{\mathrm{U}}$ denote the lower and upper bounds of scalar $x$, respectively. After some manipulations, both bounds in the second row of (36) can be written as
\begin{align}
&\bigg[\frac{\partial\mathsf{a}(\bm{t}_{m})}{\partial y_{m}}\frac{\partial\mathsf{a}(\bm{t}_{m})}{\partial z_{m}}\bigg]_{\!\mathrm{L}} \!=\! \mathsf{min} \bigg\{ [\tilde{C}_{1,m}]_{\mathrm{L}}[\tilde{C}_{2,m}]_{\mathrm{L}}, [\tilde{C}_{1,m}]_{\mathrm{L}}[\tilde{C}_{2,m}]_{\mathrm{U}}, \notag\\
&\qquad\qquad\qquad[\tilde{C}_{1,m}]_{\mathrm{U}}[\tilde{C}_{2,m}]_{\mathrm{L}}, [\tilde{C}_{1,m}]_{\mathrm{U}}[\tilde{C}_{2,m}]_{\mathrm{U}} \bigg\}, \\
&\bigg[\frac{\partial\mathsf{a}(\bm{t}_{m})}{\partial y_{m}}\frac{\partial\mathsf{a}(\bm{t}_{m})}{\partial z_{m}}\bigg]_{\!\mathrm{U}} \!=\! \mathsf{max} \bigg\{ [\tilde{C}_{1,m}]_{\mathrm{L}}[\tilde{C}_{2,m}]_{\mathrm{L}}, [\tilde{C}_{1,m}]_{\mathrm{L}}[\tilde{C}_{2,m}]_{\mathrm{U}}, \notag\\
&\qquad\qquad\qquad[\tilde{C}_{1,m}]_{\mathrm{U}}[\tilde{C}_{2,m}]_{\mathrm{L}}, [\tilde{C}_{1,m}]_{\mathrm{U}}[\tilde{C}_{2,m}]_{\mathrm{U}} \bigg\},
\end{align}
where $\tilde{C}_{1,m} \triangleq \frac{\partial\mathsf{a}(\bm{t}_{m})}{\partial y_{m}}$ and $\tilde{C}_{2,m} \triangleq \frac{\partial\mathsf{a}(\bm{t}_{m})}{\partial z_{m}}$ with
\begin{align}
&[\tilde{C}_{1,m}]_{\mathrm{L}} = -\frac{A}{2}\frac{1 - w_{0}^{2} + |w_{0}u_{0}|}{r_{0}} - w_{0}, \notag\\
&[\tilde{C}_{1,m}]_{\mathrm{U}} = \frac{A}{2}\frac{1 - w_{0}^{2} + |w_{0}u_{0}|}{r_{0}} - w_{0}, \notag\\
&[\tilde{C}_{2,m}]_{\mathrm{L}} = -\frac{A}{2}\frac{1 - u_{0}^{2} + |w_{0}u_{0}|}{r_{0}} - u_{0}, \notag\\
&[\tilde{C}_{2,m}]_{\mathrm{U}} = \frac{A}{2}\frac{1 - u_{0}^{2} + |w_{0}u_{0}|}{r_{0}} - u_{0}.
\end{align}
As suggested by (36)-(39), the right hand side of (36) is indeed independent of $\bm{t}_{m}$, which indicates that it is a global upper bound of $|H_{2,m,l}|$. Hence, by substituting (36) into (35), the global lower bound of $\mathbf{H}_{2,m,l}$ can be obtained, and the derivation of $\mathbf{M}_{m,l}$ is thus completed.

\subsection*{C. Calculation of $\nabla_{f}(\mathsf{k}(\bm{t}_{m}, \bm{\alpha}_{m};f))$}

The gradient $\nabla_{f}(\mathsf{k}(\bm{t}_{m}, \bm{\alpha}_{m};f))$ can be written as
\begin{align}
\nabla_{f}(\mathsf{k}(\bm{t}_{m}, \bm{\alpha}_{m};f)) \!\!&=\!\! \bm{w}^{\mathsf{H}}\frac{\partial\bm{b}(f)}{\partial f}\bm{b}^{\mathsf{H}}(f)\bm{w} \!\!+\!\! \bm{w}^{\mathsf{H}}\bm{b}(f)\bigg(\frac{\partial\bm{b}(f)}{\partial f}\bigg)^{\mathsf{H}}\bm{w} \notag\\
& = 2\mathsf{Re} \bigg\{ \bm{w}^{\mathsf{H}}\frac{\partial\bm{b}(f)}{\partial f}\bm{b}^{\mathsf{H}}(f)\bm{w} \bigg\},
\end{align}
where $\bm{w}$ and $\bm{b}(f)$ are short for $\bm{w}(\{ \bm{t}_{m} \}_{m=1}^{M}, f_{\mathrm{c}})$ and $\bm{b}(\{ \bm{t}_{m} \}_{m=1}^{M}, f)$, respectively. Besides, the $m$-th element of $\frac{\partial\bm{b}(f)}{\partial f}$ is given by

\begin{equation}
\bigg[\frac{\partial\bm{b}(f)}{\partial f}\bigg]_{m} = -j\frac{2\pi}{c}\mathsf{a}(\bm{t}_{m})[\bm{b}(f)]_{m}, \; m \in \mathcal{M}.
\end{equation}

\end{document}